\documentclass[10pt,twocolumn,letterpaper]{article}
\usepackage{amsmath}
\usepackage{titlesec}
\usepackage{enumitem}
\usepackage{amssymb}
\usepackage{subcaption}
\usepackage[font={small}]{caption}
\usepackage{wacv}
\usepackage{times}
\usepackage{epsfig}
\usepackage[accsupp]{axessibility}  
\usepackage{graphicx}
\usepackage{amsmath}
\usepackage[skip=2pt]{caption} 
\usepackage[accsupp]{axessibility}  

\def\wacvPaperID{0543} 

\wacvfinalcopy 

\ifwacvfinal
\fi

\ifwacvfinal
\usepackage[breaklinks=true,bookmarks=false]{hyperref}
\else
\usepackage[pagebackref=true,breaklinks=true,colorlinks,bookmarks=false]{hyperref}
\fi

\begin{document}

\title{TA-Net: Topology-Aware Network for Gland Segmentation}

\author{Haotian Wang, Min Xian\thanks{Corresponding author}, Aleksandar Vakanski\\
University of Idaho, Idaho, USA \\
{\tt\small \{haotianw, mxian, vakanski\}@uidaho.edu}
}


\maketitle
\thispagestyle{empty}

\begin{abstract}
Gland segmentation is a critical step to quantitatively assess the morphology of glands in histopathology image analysis. However, it is challenging to separate densely clustered glands accurately. Existing deep learning-based approaches attempted to use contour-based techniques to alleviate this issue but only achieved limited success. To address this challenge, we propose a novel topology-aware network (TA-Net) to accurately separate densely clustered and severely deformed glands. The proposed TA-Net has a multitask learning architecture and enhances the generalization of gland segmentation by learning shared representation from two tasks: instance segmentation and gland topology estimation. The proposed topology loss computes gland topology using gland skeletons and markers. It drives the network to generate segmentation results that comply with the true gland topology. We validate the proposed approach on the GlaS and CRAG datasets using three quantitative metrics, F1-score, object-level Dice coefficient, and object-level Hausdorff distance. Extensive experiments demonstrate that TA-Net achieves state-of-the-art performance on the two datasets. TA-Net outperforms other approaches in the presence of densely clustered glands. 
\end{abstract}

\begin{figure}[t]
\begin{center}
  \begin{subfigure}[b]{0.4\linewidth}
    \includegraphics[width=\linewidth]{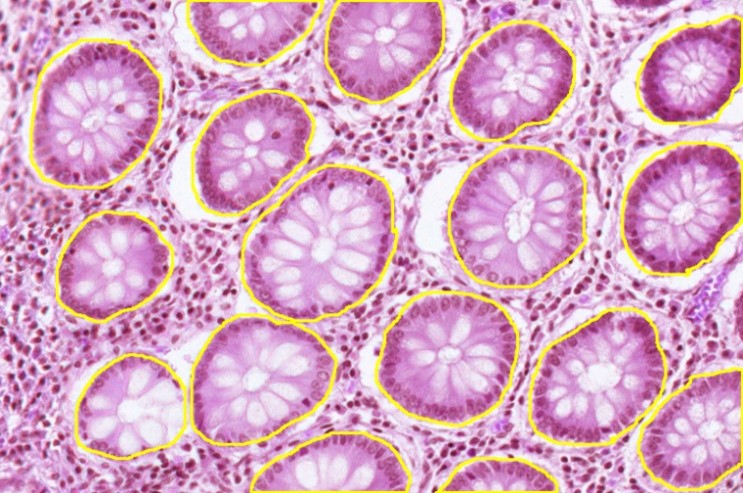}

  \end{subfigure}
  \begin{subfigure}[b]{0.4\linewidth}
    \includegraphics[width=\linewidth]{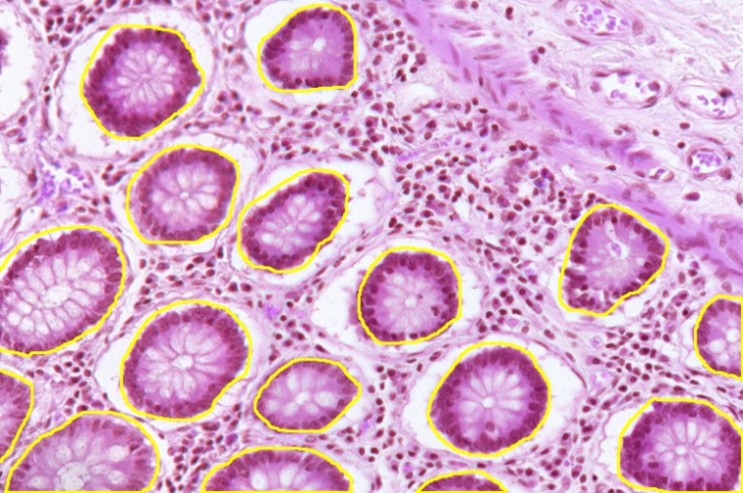}

  \end{subfigure}
  \begin{subfigure}[b]{0.4\linewidth}
    \includegraphics[width=\linewidth]{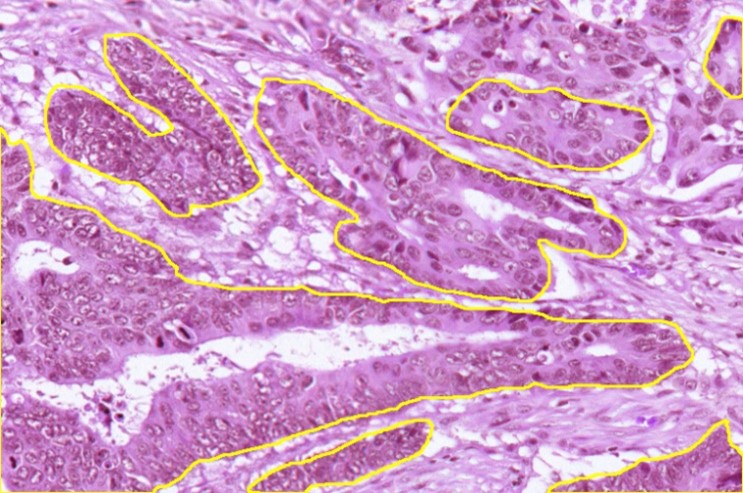}
  \end{subfigure}
  \begin{subfigure}[b]{0.4\linewidth}
    \includegraphics[width=\linewidth]{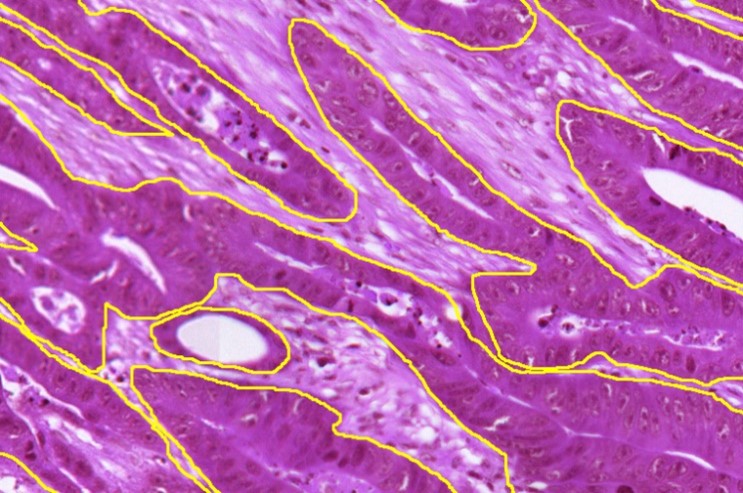}
  \end{subfigure}

\end{center}
\caption{Hematoxylin and Eosin (H\&E) stained histopathology images with labeled (yellow contours) glands. The first row shows healthy glands, the second row shows malignant glands. Noted that the malignant glands are close their neighboring glands and appear in deformed shapes.}
\label{fig:long}
\label{fig:onecol}
\end{figure}

\section{Introduction}

In histopathology image analysis,  evaluating gland morphology is crucial to determine stages of several cancers, e.g., colon cancer \cite{fleming2012colorectal}, breast cancer \cite{bloom1957histological}, and prostate cancer \cite{montironi2005gleason}. Conventionally, pathologists examine gland morphology to assess the malignancy degree using microscopes; and the whole process is time-consuming, expensive, and prone to human errors. Recently, with the availability of whole slide images (WSI), digital pathology has been achieving popularity by developing computational tools to aid routine tasks. Automatic and accurate gland segmentation is often required before calculating gland morphology. However, the task is challenging due to the large morphological differences among glands and large number of clustered glands (Figure. 1). 

Early approaches for gland segmentation focused on applying knowledge of glandular structures, e.g., morphology-based methods \cite{nguyen2010automated,paul2016gland}, and graph-based methods \cite{tosun2010graph,egger2013pcg}. These methods achieved promising performance on low-grade adenocarcinoma; but they failed in many malignant cases. Malignant glands continue to grow and invade the adjacent tissues or metastasize (Figure 1); therefore, they cluster densely and their shapes deform severely in histopthology images. Recently, deep learning-based methods provide state-of-the-art performance in many computer vision tasks \cite{badrinarayanan2017segnet,chen2017deeplab} and biomedical image analysis tasks \cite{ronneberger2015u}. Chen \emph{et~al}. \cite{chen2016dcan} proposed an FCN-based multitask learning network to generate gland regions and contours simultaneously. The complementary contour information helped separate clustered glands. Xu \emph{et~al}. \cite{xu2017gland} developed a deep three-channel network (instance, contour, and location) to jointly separate the clustered glands. Graham \emph{et~al}. \cite{graham2019mild} proposed the MILD-Net that utilized both instance and contour segmentation; and MILD-Net also involved  multi-level aggregation, atrous spatial pyramid pooling block, and dilated convolutional design. Qu \emph{et~al}. \cite{qu2019improving} proposed a full-resolutional network that outputs three-class probability maps (instance, contour and background). The strategy shared by all the above methods is to use gland contour information to separate clustered glands. However, these approaches achieved limited success. A segmentation example of densely clustered glands is shown in Figure. 2, SegNet, DCAN, and MicroNet failed to separate close glands. 

Three major challenges exist in gland segmentation using contour information: \textbf{1)} The contour strategy fails when glands are densely clustered, because overlapped glands share contour sections. Usually, in a glandular structure, epithelial nuclei form a gland border. In practice, digital WSI scanner flattens gland tissues into a near two-dimensional histopathology slice, and two or multiple glands cluster together will not have a regular and clear epithelial nuclei border. Note the red arrow in Figure. 3(a), the clustered regions do not have regular epithelial nuclei. \textbf{2)}The coarse annotations of the contours introduce noise and reduce the effectiveness of the contour strategy. A gland tissue in a 20 $\times$ magnification could be 1k $\times$ 1k pixels wide and height, and it is  difficult to annotate all contour pixels correctly. Existing datasets in the literature still have many annotation issue (Figure. 3(b)). \textbf{3)} It is difficult to identify the contours of malignant glands accurately. Because malignant glands continue to grow and become deformed. their components appear distorted. The green arrows in Figure 3(c) indicate the distorted boundary of a malignant gland.

\begin{figure}[t]
\begin{center}
  \begin{subfigure}[b]{0.3\linewidth}
    \includegraphics[width=\linewidth]{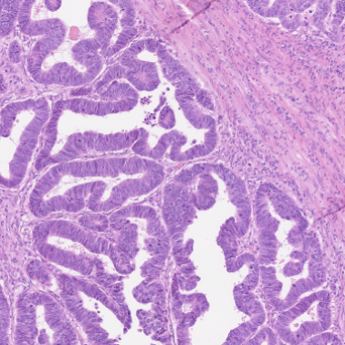}
     \caption{Image patch}
  \end{subfigure}
  \begin{subfigure}[b]{0.3\linewidth}
    \includegraphics[width=\linewidth]{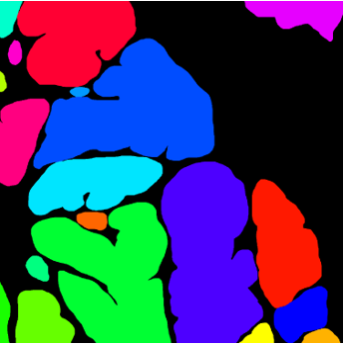}
    \caption{Ground truth}
  \end{subfigure}
  \begin{subfigure}[b]{0.3\linewidth}
    \includegraphics[width=\linewidth]{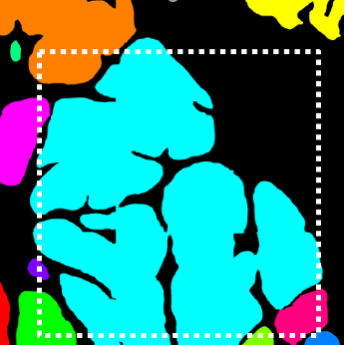}
    \caption{DCAN\cite{chen2016dcan}}
  \end{subfigure}
  \begin{subfigure}[b]{0.3\linewidth}
    \includegraphics[width=\linewidth]{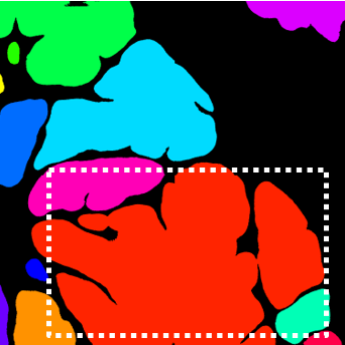}
    \caption{SegNet\cite{badrinarayanan2017segnet}}
  \end{subfigure}
  \begin{subfigure}[b]{0.3\linewidth}
    \includegraphics[width=\linewidth]{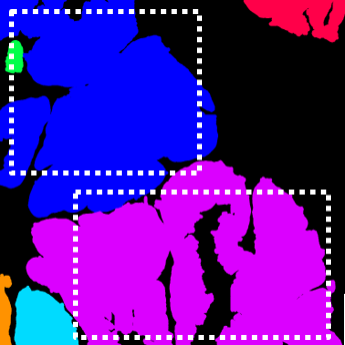}
    \caption{MicroNet\cite{raza2019micro}}
  \end{subfigure}
  \begin{subfigure}[b]{0.3\linewidth}
    \includegraphics[width=\linewidth]{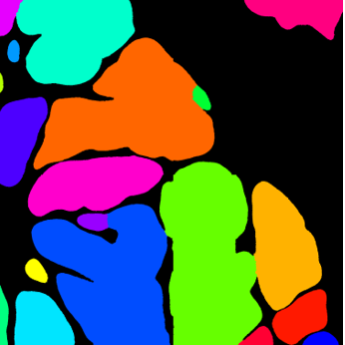}
    \caption{TA-Net}
  \end{subfigure}
\end{center}
   \caption{A segmentation example of densely clustered glands. Colors are used to differentiate different glands. The white dash rectangles highlight the poorly-separated glands.}
\label{fig:long}
\label{fig:onecol}
\end{figure}

\begin{figure}[t]
\begin{center}
  \begin{subfigure}[b]{0.9\linewidth}
    \includegraphics[width=\linewidth]{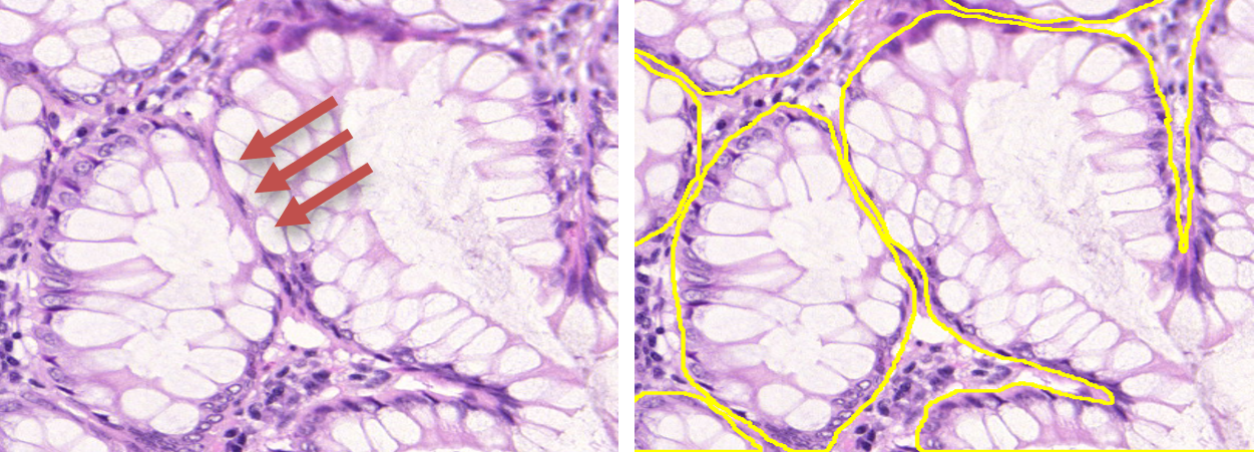}
     \caption{Lack of regular contour on densely clustered glands}
  \vspace{3px}
  \end{subfigure}
  \begin{subfigure}[b]{0.9\linewidth}
    \includegraphics[width=\linewidth]{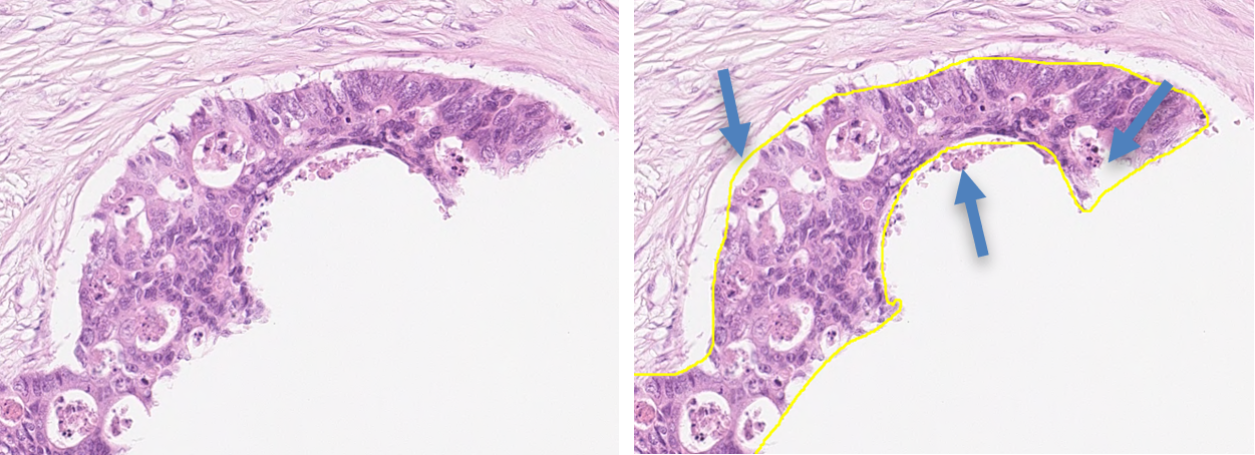}
    \caption{coarsely annotation}
  \vspace{3px}
  \end{subfigure}
  \begin{subfigure}[b]{0.9\linewidth}
    \includegraphics[width=\linewidth]{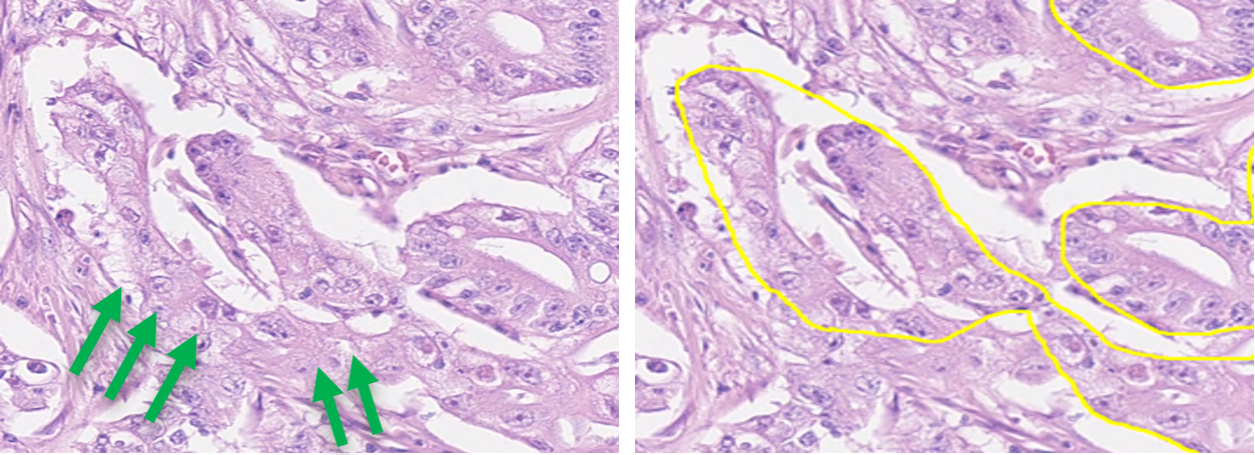}
    \caption{lack of texture and color information on malignant glands}
  \end{subfigure}

\end{center}
\caption{Issues of segmenting densely clustered glands by using a gland contour annotation. Left column shows the histopathology image patches; right column shows the image patches with the labeled gland contours (yellow).}
\label{fig:long}
\label{fig:onecol}
\end{figure}

In this paper, we introduce a new strategy that utilizes gland topology to separate densely clustered glands. The gland topology is characterized by the gland's topological skeleton, which differentiates clustered glands better than gland contours. Furthermore, the gland topology is more reliable than contours in the presence of noisy annotations. This work has two major contributions: 1) we propose a topology-aware network that learns shared representation for simultaneously instance segmentation and gland topology estimation. The topology branch of the network predicts the Medial Axis \cite{blum1967transformation} distance map to describe gland topology.  2) We propose a new topology loss by using the Medial Axis distance map and gland markers. The loss penalizes the topology difference between segmented glands and the true glands, which forces the network to generate segmentation results that adhere to the gland topology. 

\section{Related Work}

\subsection{Histopathology Gland Segmentation} 
Histopathology gland segmentation aims to segment the gland tissue from the Hematoxylin \& Eosin (H\&E) stained histopathology image. Recently, deep learning (DL)-based method successfully demonstrate the robustness and efficiency in the literature. Raza \emph{et~al}. \cite{raza2019micro} proposed the MicroNet that inputs the multiple resolutions of images patches at different down-sampling stages for better localization and context information and back-propagates the results by using multi-resolution outputs. Ding \emph{et~al}. \cite{ding2020multi} proposed a multi-scale Fully convolutional network to extract different receptive field features at different convolutional layers. These studies, as well as those described in the previous section \cite{chen2016dcan, xu2017gland, graham2019mild}, build up a deep architecture for segmenting the gland instance and contours. Yan \emph{et~al}. \cite{yan2020enabling} proposed a shape-aware adversarial learning network that integrates a deep adversarial network and a shape-preserve loss. Qu \emph{et~al}. \cite{qu2019improving} proposed a spatial loss for recognizing the glands. The proposed loss placed a spatial constraint on the boundary pixels and forced the network to learn gland shapes.

\subsection{Topology Aware Networks} Different Topology aware networks have been proposed in various natural image segmentation \cite{clough2020topological,estrada2014tree, mosinska2018beyond} and biomedical image segmentation tasks \cite{NEURIPS2019_2d95666e}. Hu \emph{et~al}. \cite{NEURIPS2019_2d95666e} introduced a loss function that make the segments have the same Betti number as the ground truths for the topological correctness; and the method utilized the topology information to make the corrections on some biological structures, e.g., broken connection. Clough \emph{et~al}. \cite{clough2020topological} proposed a method that integrated the differentiable properties of persistent homology into the network training process; the network extracts useful gradients even without ground-truth labels. Shit \emph{et~al}. \cite{shit2021cldice} introduced a centerlineDice that measured the topological similarity of the segmentation masks and their skeleta. Mosinska \emph{et~al}. \cite{mosinska2018beyond} constructed a loss that models higher-order topological information. These methods employed the regional topological constraints, e.g., the connectivity and loop-freeness. However, these methods are hard to generalized to other objects without linear structures, e.g., blood vessels, retinas, cracks, and roots. In this work, we introduce a new topology-aware network to preserve the topological skeletons of glands, which can be easily reproduced to objects with irregular shapes ans the overlapping issue. 

\begin{figure*}[t]
\begin{center}
   \includegraphics[width=0.8\linewidth]{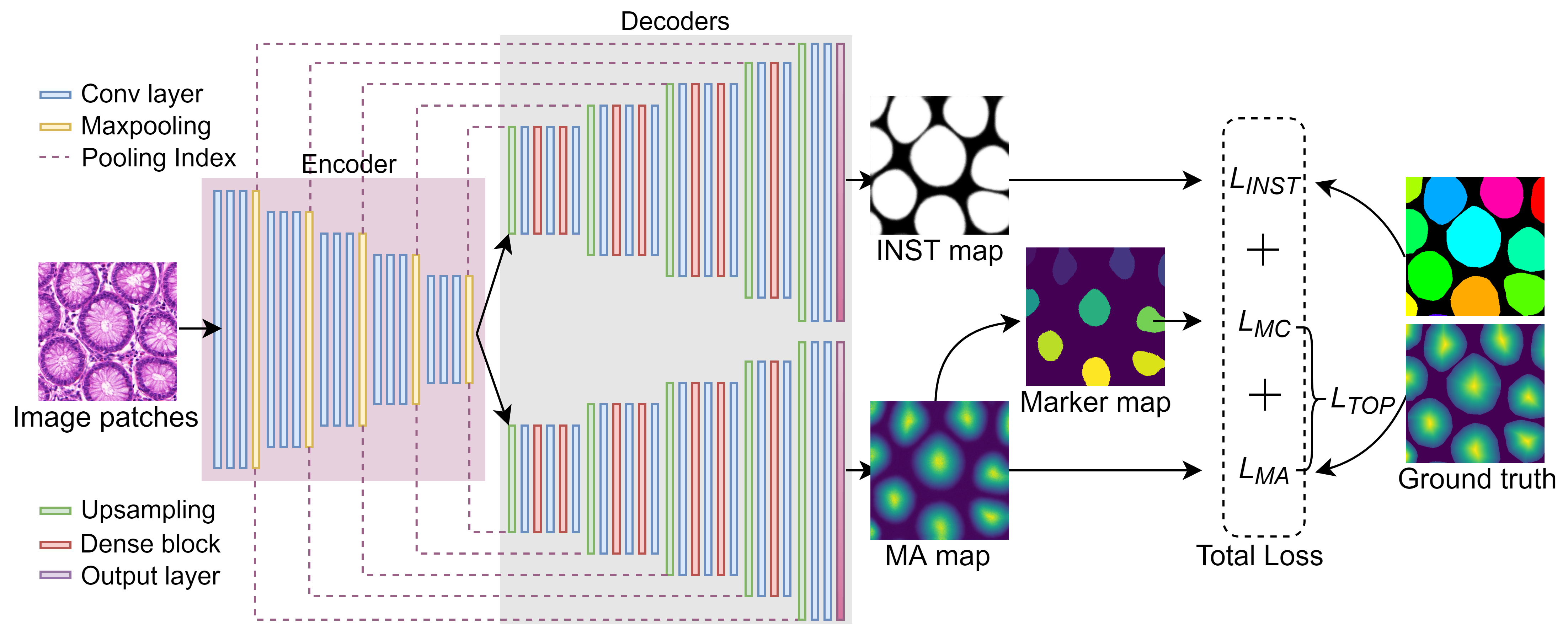}
\end{center}
   \caption{The architecture of the proposed network. The architecture takes image patches as input and outputs the gland instance (INST) map and medial axis (MA) distance map. The marker map is generated using the MA map.}
\end{figure*}

\subsection{Medial Axis (MA) Transform} The MA/topological skeleton of an object is the set of all inner points that have more than one closest point on the object's boundary. The MA transformation was firstly introduced in \cite{blum1967transformation} for shape recognition; and it is well-known as the locus of local maxima with distance transformation. The MA transform is a powerful tool for shape abstraction, and provides a shape representation that preserves the topological property of skeleton structures; these property are invariant to crop, rotation and articulation, and robust to the overlapped and clustered objects. Recently, many studies applied medial axis transforms in different computer vision tasks, e.g. segmentation \cite{noble2011atlas}, shape matching \cite{siddiqi1999shock}, recognition \cite{sebastian2001recognition}, image reconstruction \cite{tsogkas2017amat}, and body tracking \cite{shotton2012efficient}. In a densely clustered gland region, different glands may contain different skeletons. The skeleton structure may assist to distinguish the densely clustered glands. However, to the best of our knowledge, no studies have employed MA transform in histopathology image analysis.


\section{Topology-Aware Network for Gland Segmentation}

Figure. 4 illustrates the architecture of the proposed TA-Net. It is a deep multitask neural network and aims to enhance gland segmentation by learning shared representation from two tasks: gland instance (INST) segmentation and gland topology (TOP) estimation. The first task extracts glands from the background. The second task learns gland topology to separate clustered glands. 

\subsection{TA-Net Architecture}

The proposed TA-Net has one encoder and two decoder branches. The first decoder predicts foreground map for glands (INST); and the second decoder learns the topology information of glands (TOP). Two decoders share the same feature maps from the encoder. SegNet \cite{badrinarayanan2017segnet} is utilized as the backbone architecture due to its state-of-the-art performance on gland segmentation \cite{graham2019mild,sirinukunwattana2017gland} comparing with the existing benchmark networks, e.g., U-Net \cite{ronneberger2015u}, FCN8 \cite{long2015fully}, and DeepLab \cite{chen2017deeplab}. Meanwhile, dense-connected blocks \cite{huang2017densely} are applied to the decoders to ensure the large receptive field for detecting the instances over wider areas in images. In the proposed network, the two decoders have the same architectures except for the final output layers. The INST decoder ends with a 2 by 2 convolutional layer and follows a softmax activation layer; and the TOP decoder ends with a 1 by 1 convolutional layer for outputting the meidal axis distance map. 

In the encoder, three convolutional layers and the following maxpooling layer forms a downsampling convolutional block. In total, there are five downsampling blocks. In the decoder, there are five upsampling blocks that contains different numbers of stacked densely connected layers and convolutional layers. Different from the standard SegNet encoder architecture, there has three convolutional layers in the first two downsampling blocks aim at extracting more fundamental features. In TA-Net, all convolutional kernels are 3×3, and the numbers of kernels of for the convolutional layers in each block are the same. The numbers of convolutional kernels from blocks 1 to 5 in the encoder are 64, 128, 256, 512, and 512, respectively. In the two decoders, the numbers of kernels from blocks 1 to 5 are 512, 512, 256, 128, and 64, respectively; and the numbers of stacked densely-connected layers are 8, 8, 8, 8, 4, 4, and 4, respectively.

\textbf{The loss function of TA-Net.} As shown in Figure. 4, the proposed TA-Net's loss function has two terms: the instance loss ($L_{INST}$) and the topology loss ($L_{TOP}$). The total is defined by
\begin{equation}
L_{TA-Net}=L_{INST}+\alpha \cdot L_{TOP}
\end{equation}
where $\alpha$ denotes the contribution of the topology loss. The loss $L_{INST}$ is the cross-entropy (CE) loss on gland instances map for segmenting the foreground gland instances from the background. The loss $L_{TOP}$ is discussed in Section 2.2, $\alpha$ controls the contribution of the $L_{TOP}$ loss. 

\subsection{Topology Loss}
The proposed topology loss is given by
\begin{equation}
L_{TOP}=L_{MA}+L_{MC}
\end{equation}
where $L_{MA}$ is computed using the medial axis distance map to preserve the geometry of glands, and $L_{MC}$ uses markers in glands to avoid over-segmentation and under-segmentation. 

\textbf{Medial Axis (MA) Distance Map.} The MA/topological skeleton of an object is the set of all inner points that have more than one closest point on the object's boundary. It is well-known as the locus of local maxima with distance transformation. In this work, we employ the MA-based distance map to model the topological property of glands. 
Let $G=\{G_{i}\}_{i=1}^n$ be a set of glands in an image patch, and \textit{n} be the number of glands. For every gland from $G_{1}$ to $G_{n}$, the MA transformation iterates the one-pixel morphological erosion process starting through the gland contour. The topological skeleton of a gland (Fig. 5(c)) is a set of points having more than one closest point on a gland's contour; and the skeletons are one-pixel width and follow the same connectivity as the original gland shape. The number of iterations from a gland's contour to the topological skeleton is normalized to form the MA distance map. The MA distance map value at point $p_{j}$ is defined by 
\begin{equation}
\begin{split}
MA(p_{j})=\hspace{2.5in} \\
\begin{cases}
\frac{d(p_{j})}{\max_{\forall p_{k} \in G_{i}}\{d(p_{k})\} - \min_{\forall p_{k} \in G_{i}} \{d(p_{k})\}}, &if \exists G_{i} \in G, p_{j} \in G_{i}\\
0, &\text{otherwise}
\end{cases}
\end{split}
\end{equation}
where $d(p_{j})$ is the number of erosion iterations from point $p_{j}$ to the corresponding skeleton. 

\begin{figure}[t]
\begin{center}
  \begin{subfigure}[b]{0.24\linewidth}
    \includegraphics[width=\linewidth]{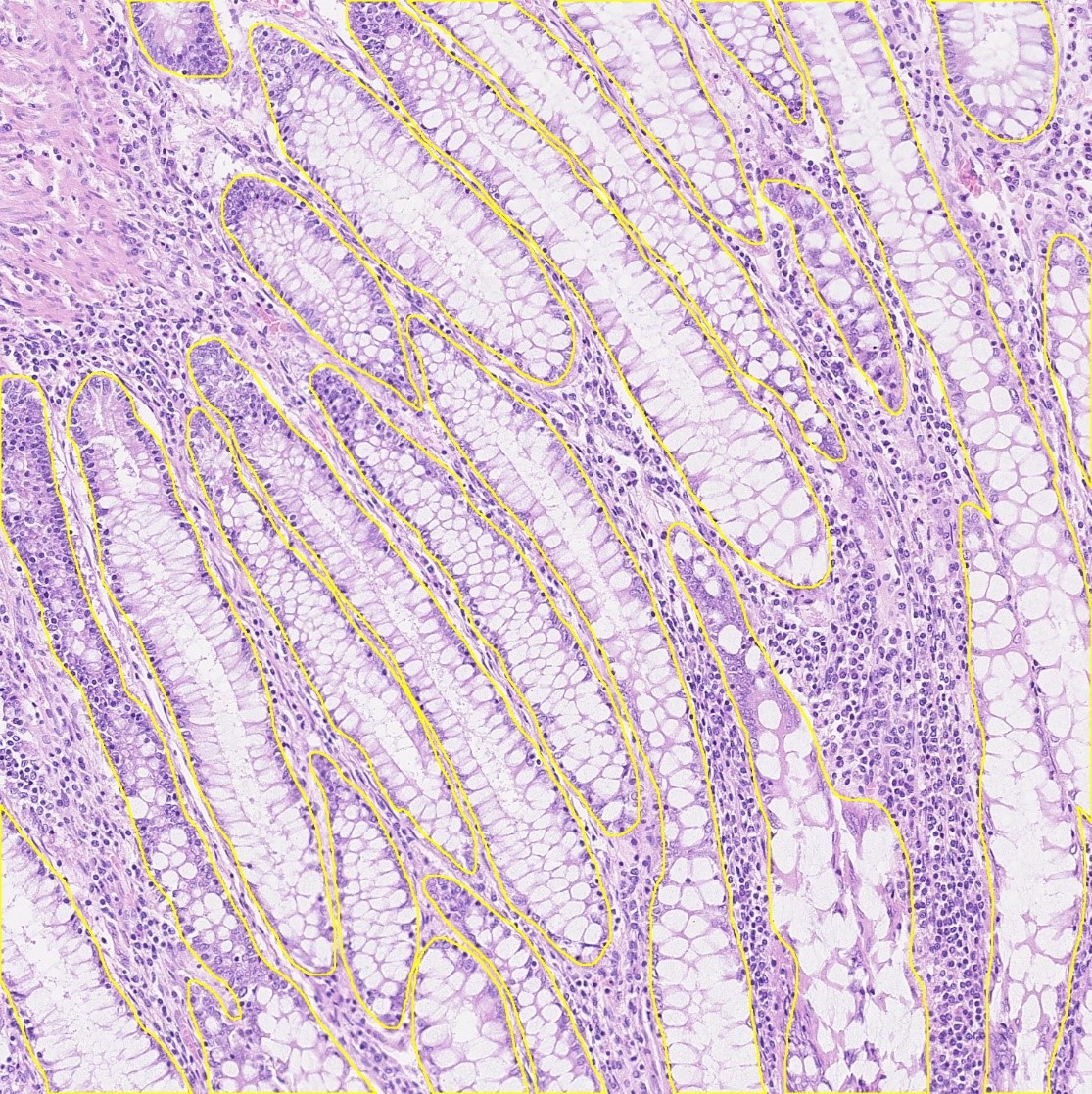}
  \end{subfigure}
  \begin{subfigure}[b]{0.24\linewidth}
    \includegraphics[width=\linewidth]{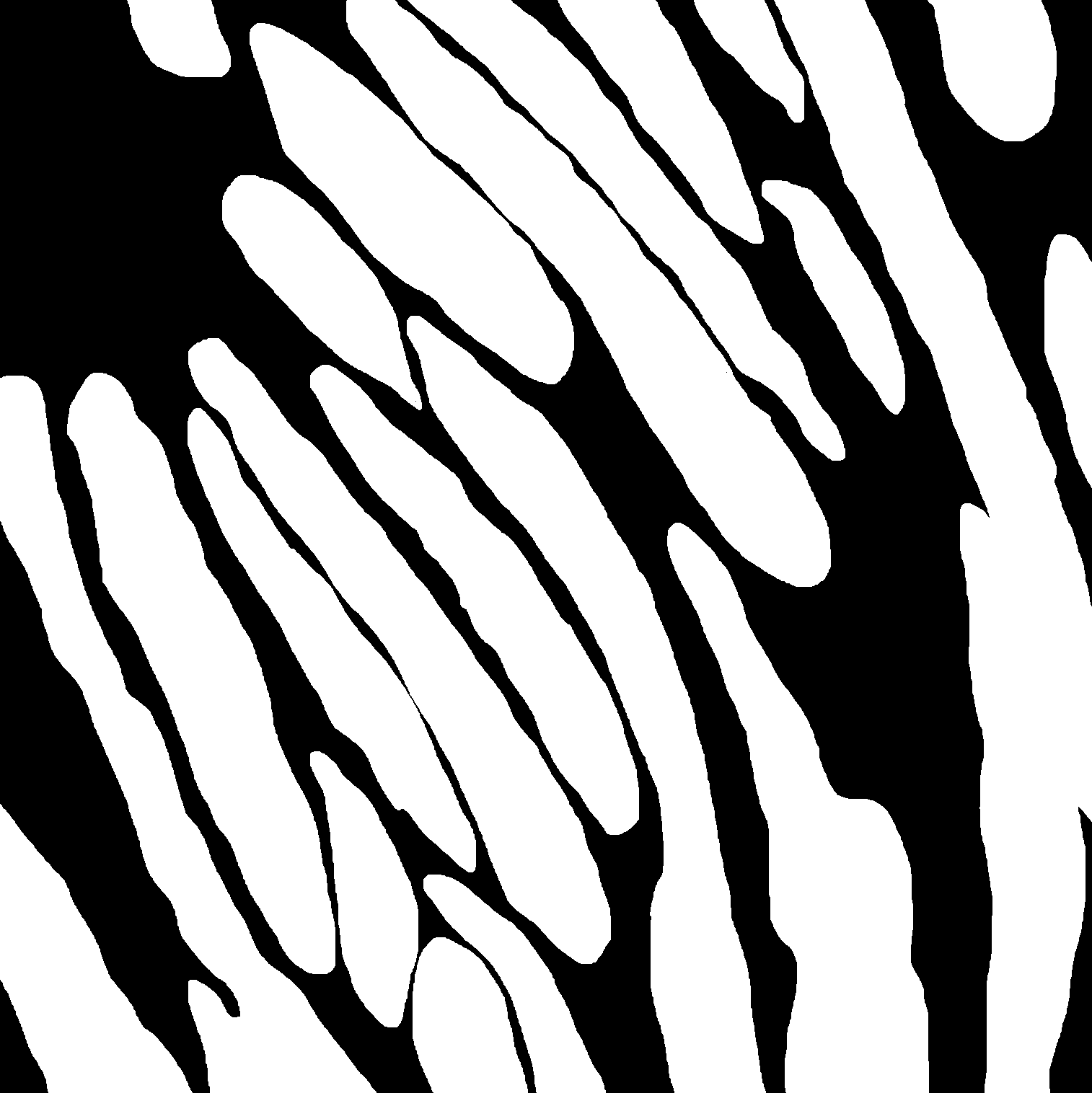}
  \end{subfigure}
  \begin{subfigure}[b]{0.24\linewidth}
    \includegraphics[width=\linewidth]{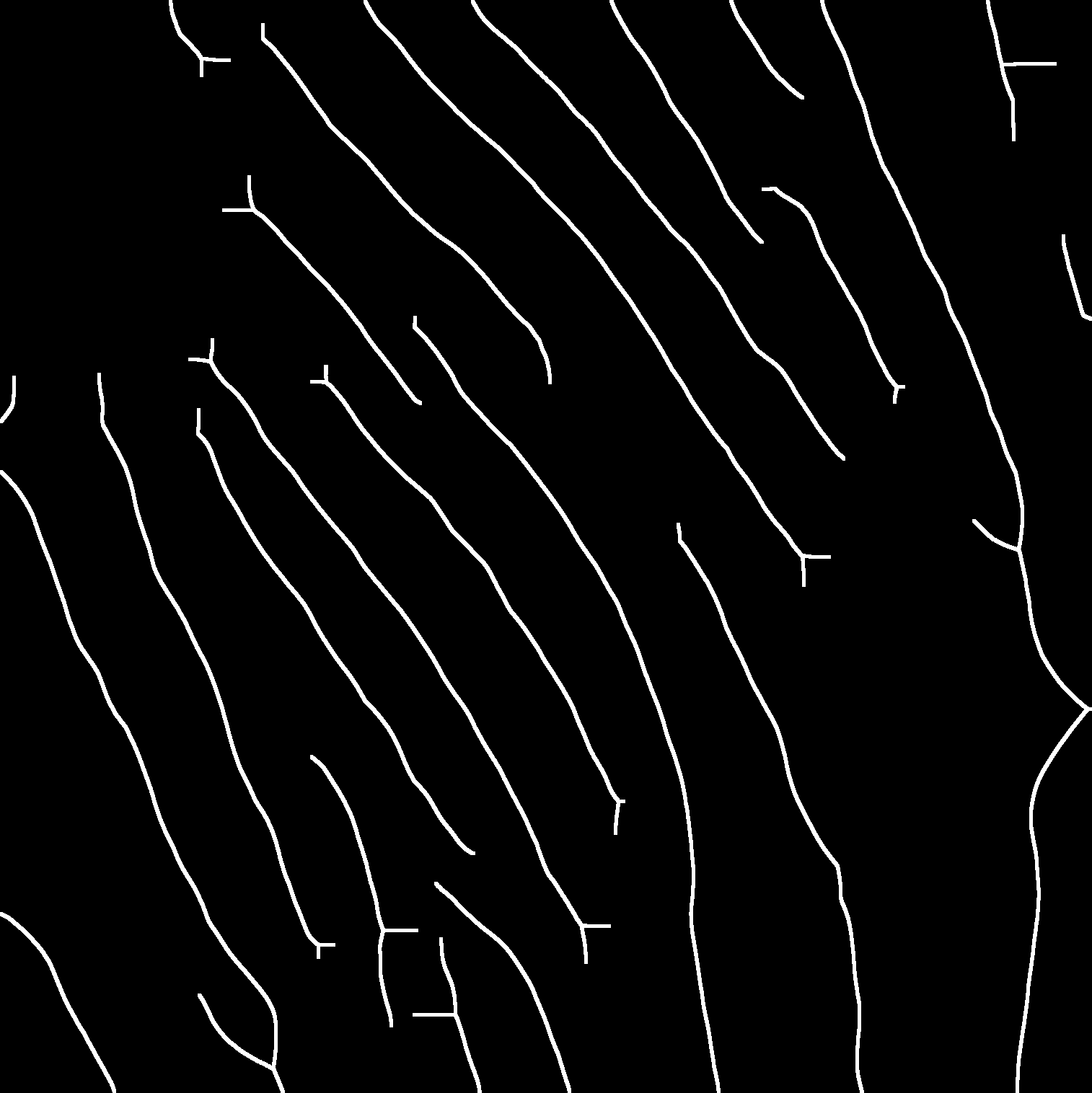}
  \end{subfigure}
  \begin{subfigure}[b]{0.24\linewidth}
    \includegraphics[width=\linewidth]{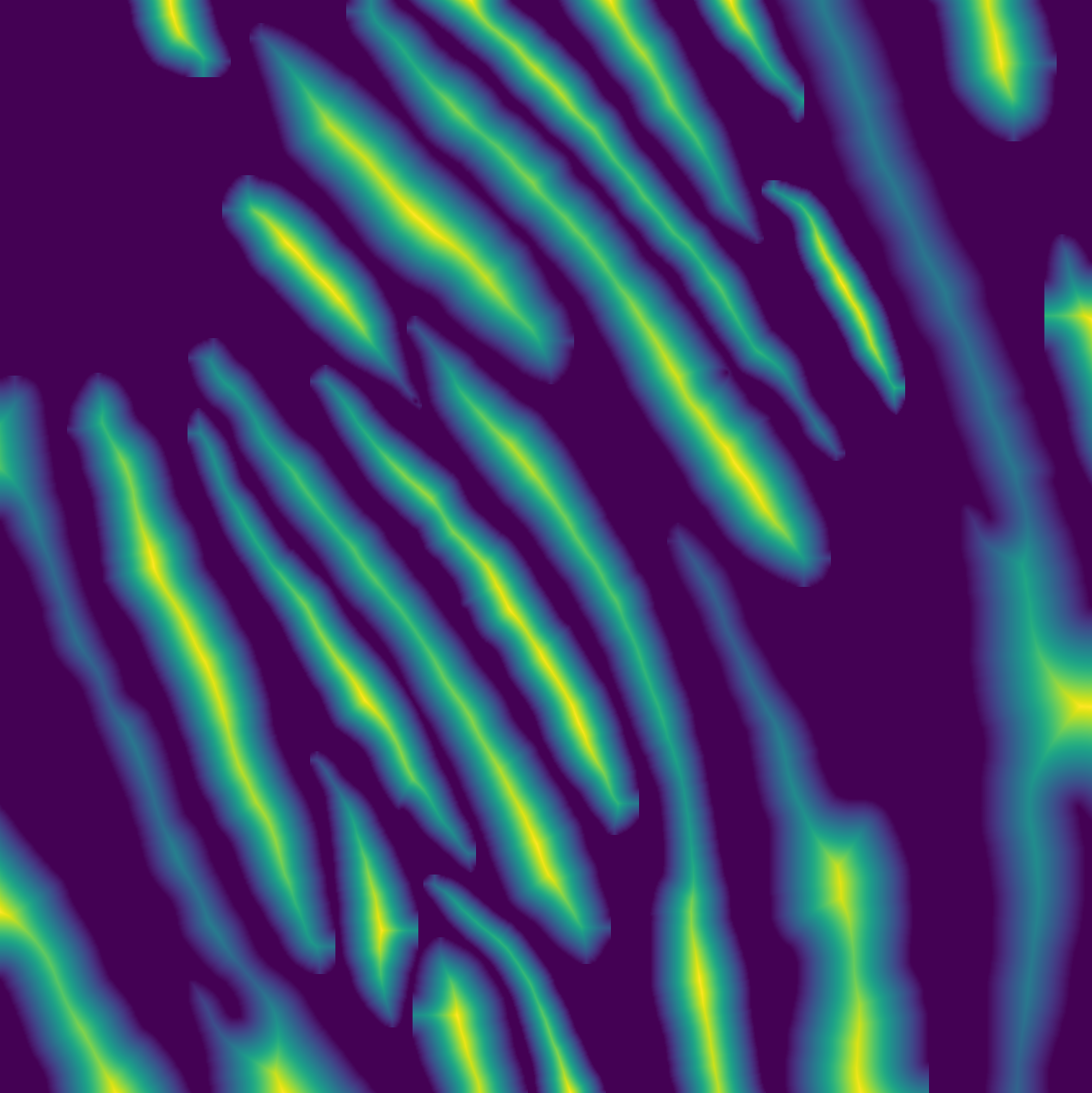}
  \end{subfigure}
  \begin{subfigure}[b]{0.24\linewidth}
    \includegraphics[width=\linewidth]{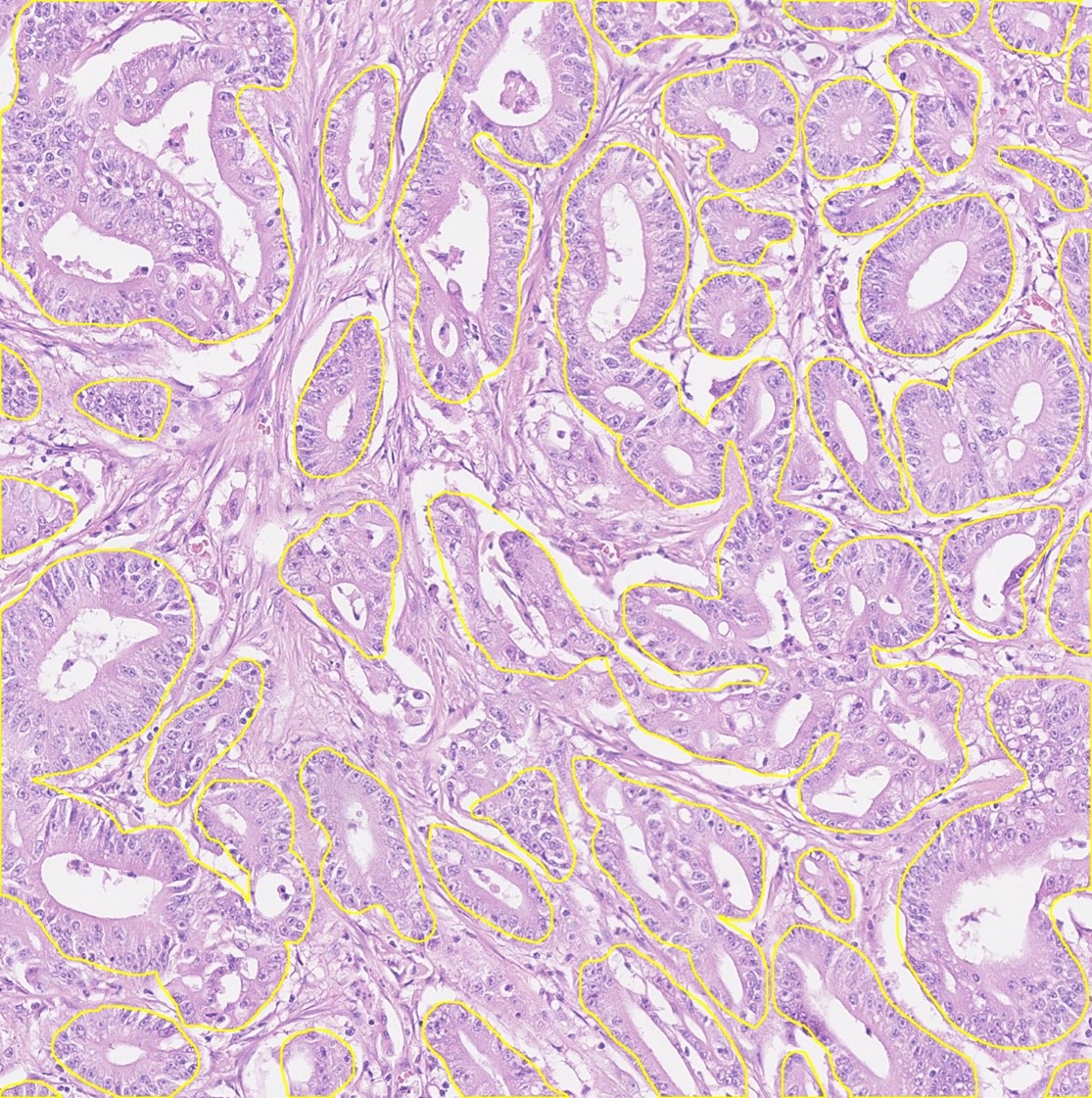}
\caption{}
  \end{subfigure}
  \begin{subfigure}[b]{0.24\linewidth}
    \includegraphics[width=\linewidth]{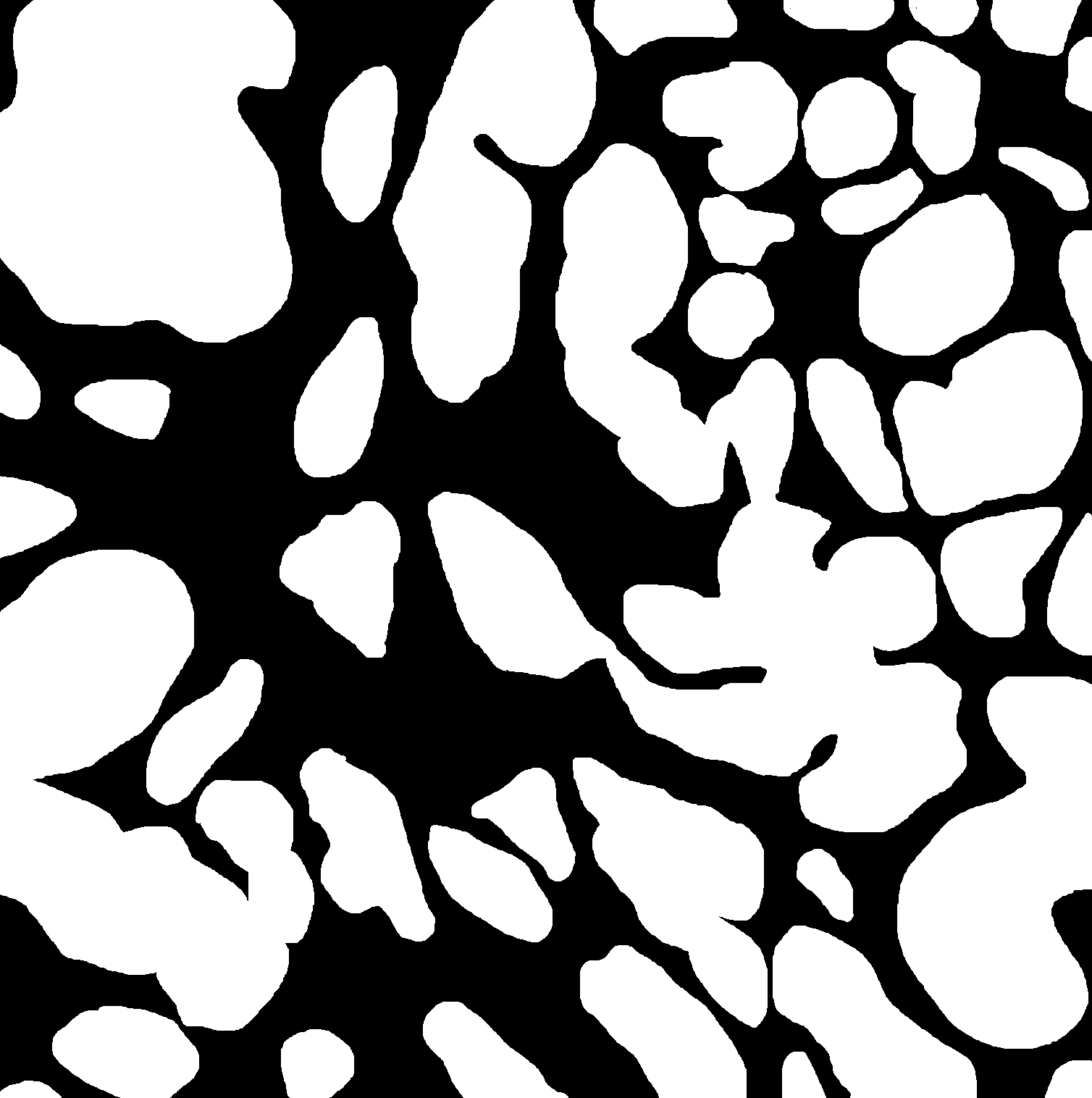}
\caption{}
  \end{subfigure}
  \begin{subfigure}[b]{0.24\linewidth}
    \includegraphics[width=\linewidth]{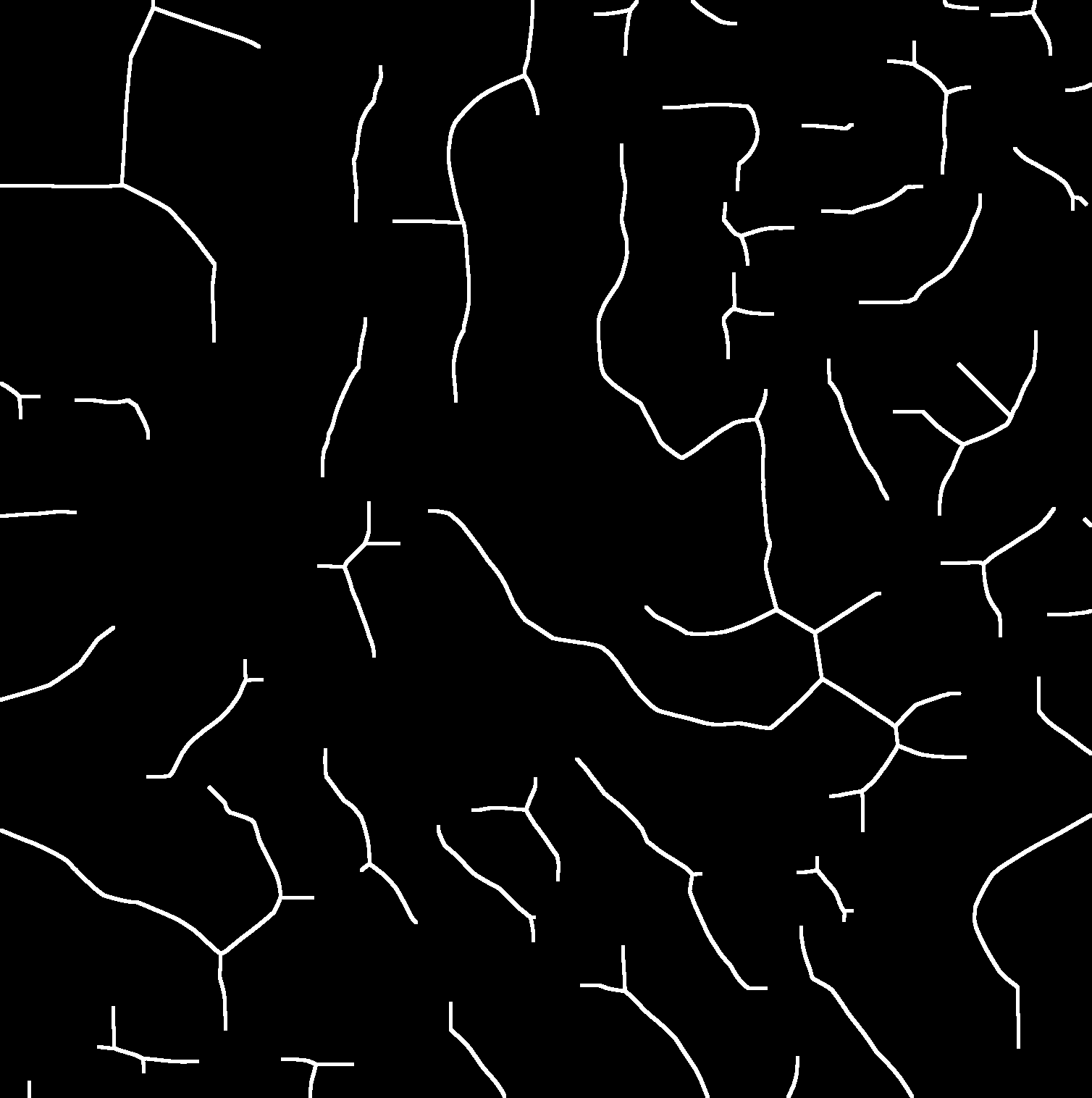}
\caption{}
  \end{subfigure}
  \begin{subfigure}[b]{0.24\linewidth}
    \includegraphics[width=\linewidth]{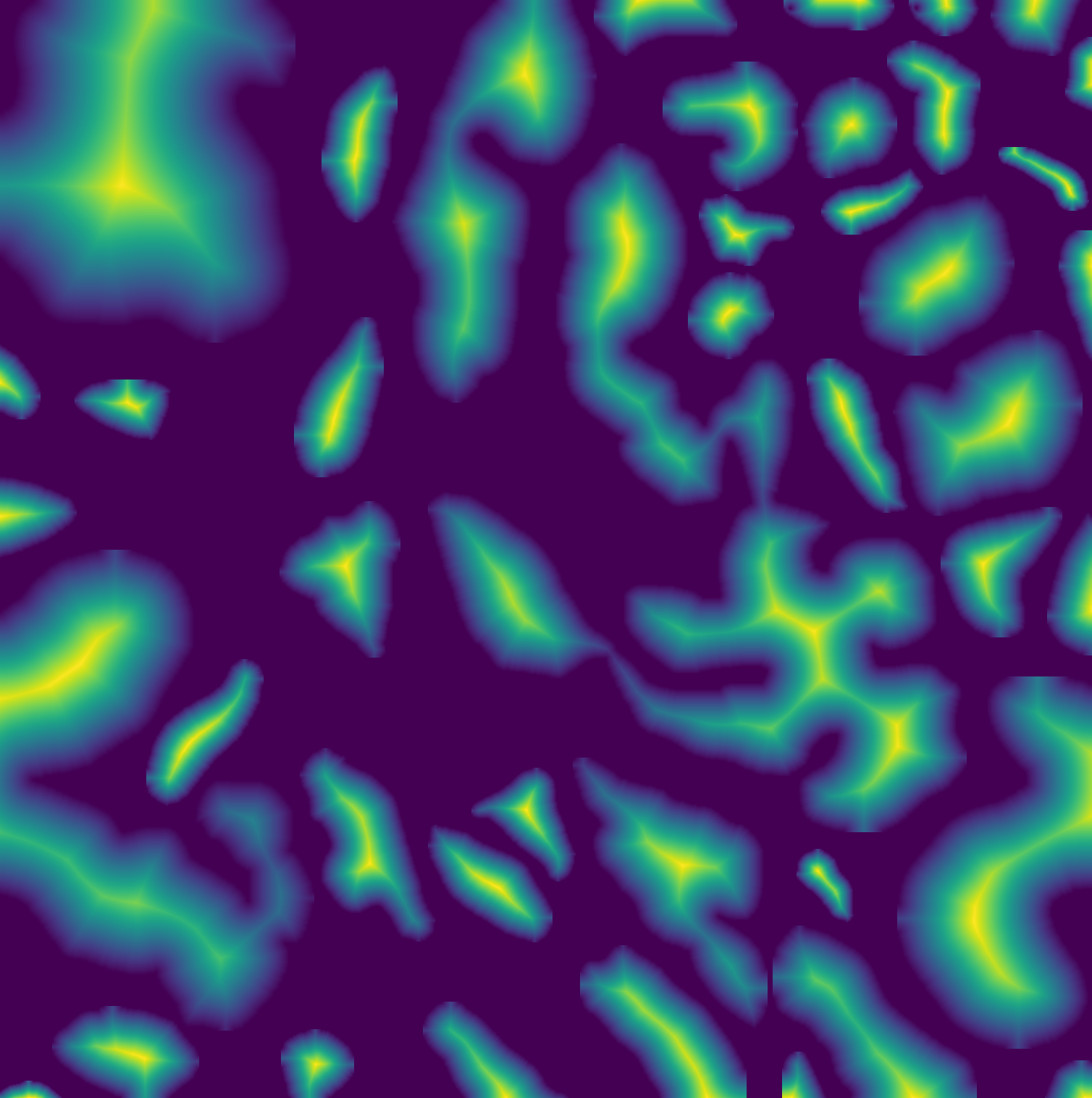}
\caption{}
  \end{subfigure}
  
\end{center}
   \caption{Examples of Medial axis (MA) transformation. a) A histopathology image patch with labeled (yellow contours) clustered glands; b) the binary annotation of gland regions; c) the topological skeletons of glands, noted that skeletons are morphological dilated to be visible; and d) the MA distance map.  }
\label{fig:long}
\label{fig:onecol}
\end{figure}

Examples of the MA distance map is shown in Figure. 5(d). As shown in Figure. 5(a), it is challenging to separate the clustered glands; however, in Figure. 5(c) and (d), the skeleton and distance map emphasize the geometrical and topological properties of each gland, and clearly separate all clustered glands. Eq.(3) and the MA transformation are applied to generate the ground truth for the MA distance map. 
To ensure that the proposed gland segmentation network preserves gland topology, we design the second decoder branch to predict the MA distance map. The loss function $L_{MA}$ is defined by
\begin{equation}
L_{MA}=\frac{1}{m} \sum_{j=1}^m (MA(p_{j})-\widehat{MA}(p_{j}))^2   
\end{equation}
where $\widehat{MA}$  denotes the predicted distance map, and \textit{m} is the number of image pixels.

\textbf{Marker loss.} The Watershed algorithm is commonly applied as a postprocessing step to produce the fine segmentation, especially separating the clustered objects. The watershed markers represent the number and locations of objects, and are critical for accurate segmentation. More markers lead to over-segmentation, and fewer markers produce under-segmented results. We introduced the marker loss to separate clustered glands and prevent over-segmentation. The marker loss is defined as the Dice loss between predicted marker map ($\widehat{MC}$) and the true marker map (\textit{MC})
\begin{equation}
L_{MC}=Dice(MC,\widehat{MC})      
\end{equation}
The predicted marker map is generated by thresholding the outputs from the medial axis distance map.

\section{Experimental Results}

\subsection{Datasets and Evaluation Metrics}

\textbf{Datasets.} The Colorectal adenocarcinoma gland (CRAG) dataset \cite{graham2019mild} and the Gland Segmentation challenge (GlaS) dataset \cite{sirinukunwattana2017gland} are used in this work. CRAG has 213 H\&E-stained histopathology images from 38 WSI images. The scanned image size is 1512 × 1516 pixels with the corresponding instance-level ground truth. The training set has 173 images and the test set has 40 images with different cancer grades. The GlaS dataset  has 165 H\&E-stained histopathology images extracted from 16 WSI images. The image size mostly is 775 × 522 pixels. The training set has 85 images (37 benign and 48 malignant). The test set is split into two sets: Test A (60 images) and Test B (20 images), because two test sets are releases in different stages in GlaS challenge. Both datasets are scanned with a 20× objective magnification. The CRAG dataset has more densely clustered glands.

\textbf{Evaluation Metrics.} We use the F1-score, object-level Dice coefficient (Obj-D), and object-level Hausdorff distance (Obj-H). In the F1 score, a segmented gland is counted as a true positive if it has \textgreater50\% overlap with the ground truth, and counted as a false positive (FP) if otherwise; and all missed glands in the ground truth are counted as false negatives. Refer to \cite{sirinukunwattana2017gland} for detailed descriptions of the Obj-D and Obj-H in Colon Histology Images Challenge Contest (GlaS) at MICCAI 2015. 

\begin{table*}[t]
\small
\begin{center}
\begin{tabular}{|l|l|c|c|c|c|}
\hline Datasets & Methods & Year & F1(\%)$\uparrow$ & Obj-D(\%)$\uparrow$ & Obj-H$\downarrow$ \\
\hline\hline
GlaS & SegNet\cite{badrinarayanan2017segnet} & 2016 & 83.1 & 84.9 & 76.6\\
                & DCAN\cite{chen2016dcan} & 2016 & 81.4 & 83.9 & 102.9\\
                & DeepLab\cite{chen2017deeplab} & 2017 & 83.7 & 84.5 & 80.5\\
                & MILD-Net\cite{graham2019mild} & 2019 & 87.9 & 87.5 & 73.7\\
                & Micro-Net\cite{raza2019micro} & 2019 & 86.5 & 87.6 & 70.4\\
                & FullNet\cite{qu2019improving} & 2019 & 88.9 & 88.5 & 63.0\\
                & DSE\cite{xie2019deep} & 2019 & 89.4 & 89.9 & 55.9\\
                & MSFCN\cite{ding2020multi} & 2020 & 89.3 & 89.9 & 53.1\\    
                & Yan \emph{et~al}.\cite{yan2020enabling} & 2020 & \textbf{90.7} & 89.3 & 58.7\\       
                & TA-Net & 2021 & 90.5 & \textbf{90.2} & \textbf{50.8} \\
\hline
CRAG & SegNet\cite{badrinarayanan2017segnet} & 2016 & 77.4 & 85.3 & 134.7\\
                & DCAN\cite{chen2016dcan} & 2016 & 73.6 & 79.4 & 218.8\\
                & DeepLab\cite{chen2017deeplab} & 2017 & 64.8 & 74.5 & 281.4\\
                & MILD-Net\cite{graham2019mild} & 2019 & 82.5 & 87.5 & 160.1\\
                & DSE\cite{xie2019deep} & 2019 & 83.5 & 88.9 & 120.1\\
                & MSFCN\cite{ding2020multi} & 2020 & 82.5 & 89.2 & 130.4\\    
                & TA-Net & 2021 & \textbf{84.2} & \textbf{89.3} & \textbf{105.2} \\
\hline
\end{tabular}
\end{center}
\caption{Overall segmentation performance on GlaS and CRAG datasets.}
\end{table*}

\subsection{Implementation Details }

The TA-Net is trained and tested using a deep learning server with an NVIDIA Quadro RTX 8000 GPU, 512 GB memory, and two 2.4 GHz Intel Xeon 4210R CPUs.

The patch size of the GlaS dataset is 512 × 512 pixels, and the patch size of the CRAG dataset is 768 × 768 pixels. Different patch sizes are applied because 1) the two datasets have images with different sizes; and 2) larger patches improve the performance on segmenting white lumen regions inside the gland tissues. The larger patches could generate the whole white lumen region in the gland in one patch. The GlaS dataset generates 340 training patches and 320 test patches. The CRAG dataset generates 692 training patches and 160 test patches. The augmentation approaches, e.g., random flip, random rotation, Gaussian blur, and median blur, are utilized in the training stage. The segmentation results of image patches are merged to form images of the same size as the original images. 

The training epoch is set as 200, and the initial learning rate for the Adam optimizer is set as 10-4 and is reduced to 10-5 after 100 epochs. The batch size is 4 for training the model.

The postprocessing applies the Watershed algorithm to produce the final output. We apply a threshold value to the outputs from the instance branch (INST) to generate the glands binary map, and utilize it as the Watershed filling region. The output from the MA distance branch are the local elevation of those glands. Further, thersholding the MA distance map to generate the Watershed markers. Morphology operations, e.g., fill the holes, remove the small objects are utilized to generate a fine glands regions and makers. In the end, the generated gland region, gland elevation and markers are input to Watershed algorithm for fine gland segmentation results.

\subsection{Results and Discussion}

In this section, we discuss the overall performance of our method, followed by the results using the contour map,  and single-task/multitask networks. Finally, we discuss the performance of our network on different distance-metrics. 

\vspace*{3mm}
\textbf{Overall performance.} We compare the proposed method with nine recently published approaches using the GlaS dataset and six approaches using the CRAG datasets. We implemented \cite{badrinarayanan2017segnet, chen2017deeplab, chen2016dcan} by following the same strategies in the original papers and the rest of other results are cited from their original papers. The test performance of GlaS dataset is reported as the average performance on its test A and test B sets. We employed F1-score, object-level Dice coefficient (Obj-D), and object-level Hausdroff (Obj-H) distance to measure the overall performance.

Table 1 shows the test performance of different approaches on two public datasets. The proposed TA-Net outperforms all other methods on CRAG datasets in terms of the F1 score, Obj-D; and achieved the best results in Obj-D, Obj-H and the second-best results in F1-score on GlaS datasets. The GlaS dataset has a small number of densely clustered gland regions; therefore, comparing with the second-highest result (Yan \emph{et~al}. \cite{ding2020multi}), TA-Net is only slightly better, e.g. 4.3\% improvement of the Obj-H.  The CRAG dataset has more densely clustered glands, and our method improves the Obj-H significantly (19.3\%). Figure. 6(d) demonstrates that TA-Net separates densely clustered glands accurately.  

\begin{figure*}
\begin{center}
  \begin{subfigure}[b]{0.18\linewidth}
    \includegraphics[width=\linewidth]{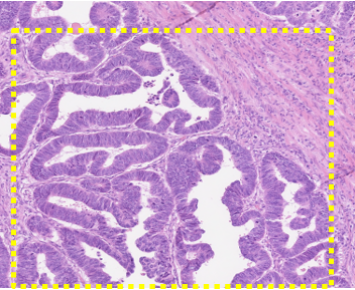}
  \end{subfigure}
  \begin{subfigure}[b]{0.18\linewidth}
    \includegraphics[width=\linewidth]{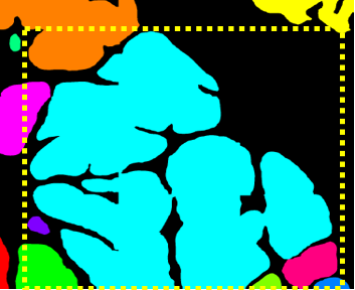}
  \end{subfigure}
  \begin{subfigure}[b]{0.18\linewidth}
    \includegraphics[width=\linewidth]{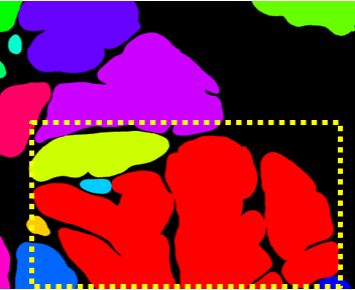}
  \end{subfigure}
  \begin{subfigure}[b]{0.18\linewidth}
    \includegraphics[width=\linewidth]{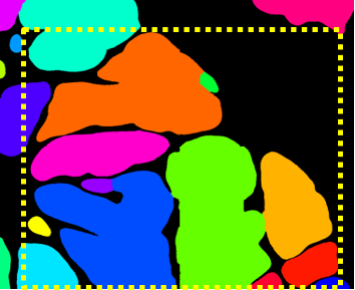}
  \end{subfigure}
  \begin{subfigure}[b]{0.18\linewidth}
    \includegraphics[width=\linewidth]{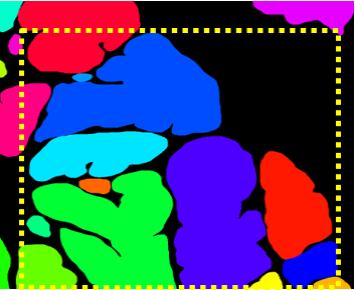}
  \end{subfigure}
  
  \begin{subfigure}[b]{0.18\linewidth}
    \includegraphics[width=\linewidth]{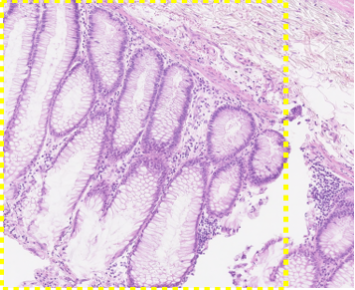}
  \end{subfigure}
  \begin{subfigure}[b]{0.18\linewidth}
    \includegraphics[width=\linewidth]{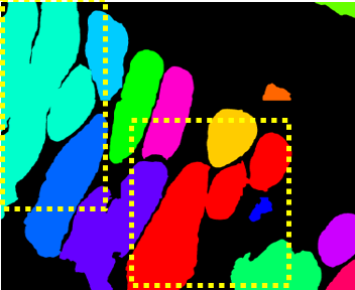}
  \end{subfigure}
  \begin{subfigure}[b]{0.18\linewidth}
    \includegraphics[width=\linewidth]{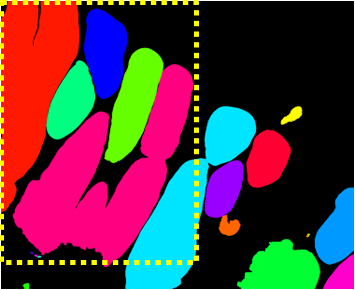}
  \end{subfigure}
  \begin{subfigure}[b]{0.18\linewidth}
    \includegraphics[width=\linewidth]{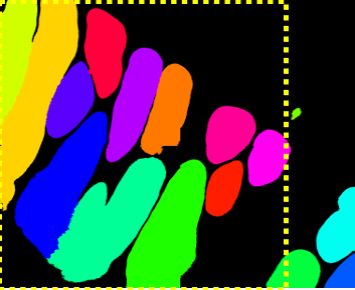}
  \end{subfigure}
  \begin{subfigure}[b]{0.18\linewidth}
    \includegraphics[width=\linewidth]{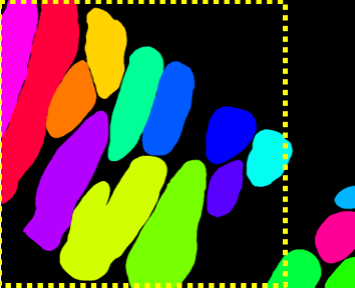}
  \end{subfigure}
  
  \begin{subfigure}[b]{0.18\linewidth}
    \includegraphics[width=\linewidth]{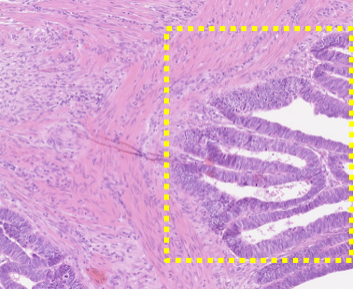}
  \end{subfigure}
  \begin{subfigure}[b]{0.18\linewidth}
    \includegraphics[width=\linewidth]{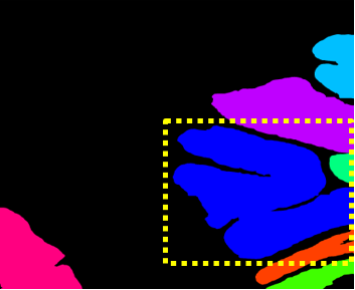}
  \end{subfigure}
  \begin{subfigure}[b]{0.18\linewidth}
    \includegraphics[width=\linewidth]{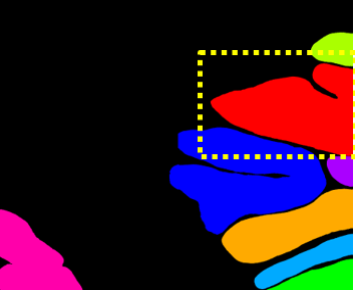}
  \end{subfigure}
  \begin{subfigure}[b]{0.18\linewidth}
    \includegraphics[width=\linewidth]{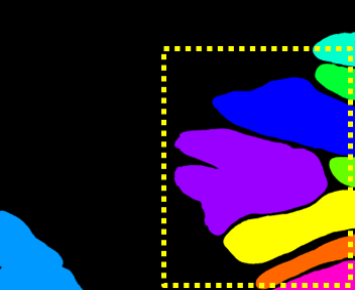}
  \end{subfigure}
  \begin{subfigure}[b]{0.18\linewidth}
    \includegraphics[width=\linewidth]{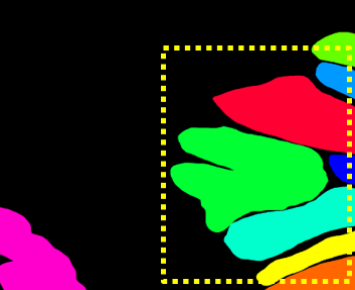}
  \end{subfigure}
  
  \begin{subfigure}[b]{0.18\linewidth}
    \includegraphics[width=\linewidth]{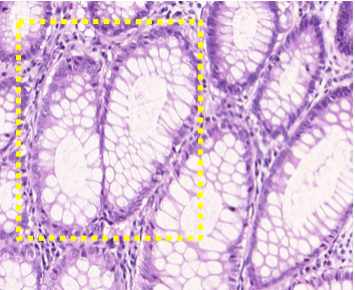}
  \end{subfigure}
  \begin{subfigure}[b]{0.18\linewidth}
    \includegraphics[width=\linewidth]{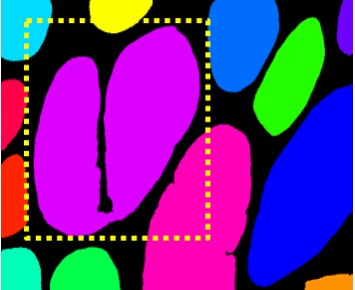}
  \end{subfigure}
  \begin{subfigure}[b]{0.18\linewidth}
    \includegraphics[width=\linewidth]{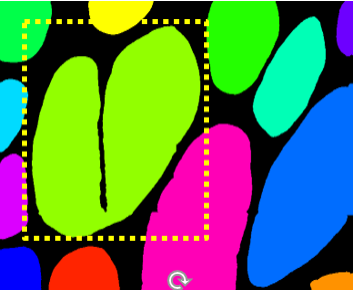}
  \end{subfigure}
  \begin{subfigure}[b]{0.18\linewidth}
    \includegraphics[width=\linewidth]{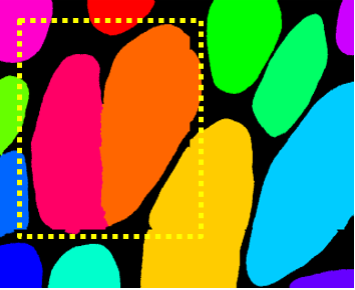}
  \end{subfigure}
  \begin{subfigure}[b]{0.18\linewidth}
    \includegraphics[width=\linewidth]{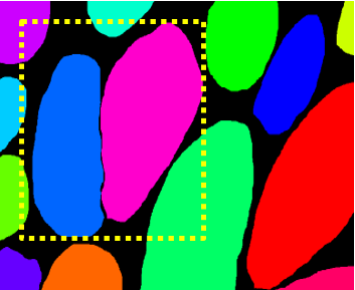}
  \end{subfigure}
  
  \begin{subfigure}[b]{0.18\linewidth}
    \includegraphics[width=\linewidth]{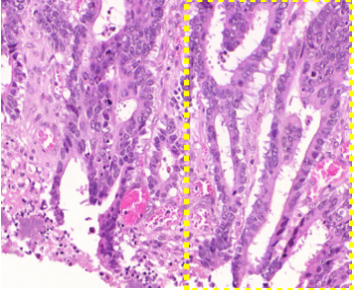}
     \caption{Original Images}
  \end{subfigure}
  \begin{subfigure}[b]{0.18\linewidth}
    \includegraphics[width=\linewidth]{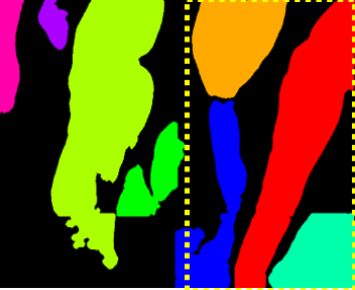}
     \caption{DCAN}
  \end{subfigure}
  \begin{subfigure}[b]{0.18\linewidth}
    \includegraphics[width=\linewidth]{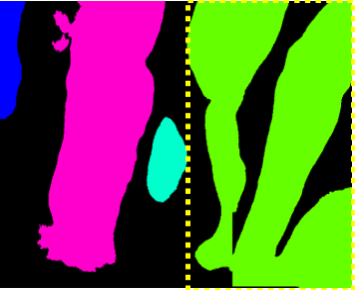}
     \caption{Ours-CNT}
  \end{subfigure}
  \begin{subfigure}[b]{0.18\linewidth}
    \includegraphics[width=\linewidth]{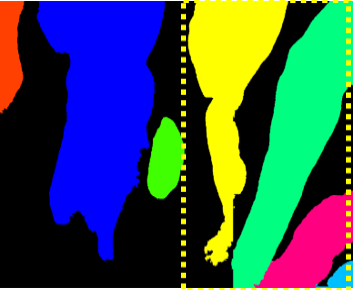}
     \caption{TA-Net}
  \end{subfigure}
  \begin{subfigure}[b]{0.18\linewidth}
    \includegraphics[width=\linewidth]{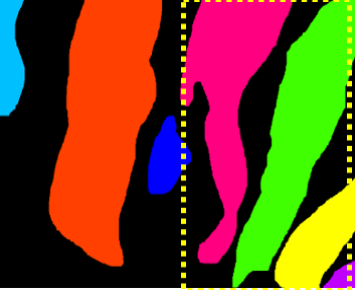}
     \caption{Ground Truth}
  \end{subfigure}
\end{center}
   \caption{Segmentation results of five image patches (top three from CRAG, bottom two from GlaS). Different colors represent different glands. Yellow dash region highlighted the clustered gland regions.}
\label{fig:long}
\label{fig:onecol}
\end{figure*}

\begin{table}[t]
\small
\begin{center}
\begin{tabular}{|l|l|c|c|c|}
\hline
Datasets & Methods & F1(\%)$\uparrow$ & Obj-D(\%)$\uparrow$ & Obj-H$\downarrow$ \\
\hline\hline
GlaS & Ours-INST & 86.4 & 88.3 & 65.4\\
                & Ours-MA & 80.2 & 84.2 & 105.8\\
                & Ours-CNT & 89.1 & 88.2 & 54.4\\
                & TA-Net & \textbf{90.5} & \textbf{90.2} & \textbf{50.8} \\
\hline
CRAG & Ours-INST & 78.9 & 86.1 & 125.6\\
                & Ours-MA & 74.8 & 80.3 & 200.8\\
                & Ours-CNT & 81.3 & 85.8 & 164.5\\
                & TA-Net & \textbf{84.2} & \textbf{89.3} & \textbf{105.2} \\
\hline
\end{tabular}
\end{center}
\caption{Ablation study on multitask learning and decoders}
\end{table}

\vspace*{3mm}
\textbf{MA distance map vs. contour map.} We compared the proposed TA-Net with a multitask network (Ours-CNT) which has the same architecture as TA-Net but outputs the gland instance map and the gland contour map. The only difference between the two networks is that TA-Net uses the MA distance map, while Ours-CNT uses the gland contour map as the ground truth of the second decoder. As shown in Table.2, TA-Net comparing to the Ours-CNT, Obj-H has been improved by 5.7\% and 35\% on the GlaS and CRAG datasets, respectively. The results demonstrate that the network using the MA distance map generates more accurate gland contours than the contour map based network, especially on the CRAG dataset. Fig. 6 demonstrates that contour map-based strategy fails to separate many clustered glands (in dashed rectangles). 

\vspace*{3mm}
\textbf{Multitask network vs. Singletask network.} In addition, we comparing the proposed multitask learning network with the single task learning network, which outputs the binary instance branch only (Ours-INST), and MA distance branch only (Ours-MA). In experiment of the Ours-INST, the Watershed algorithm is applied in the postprocessing for separating the clustered glands. In experiment of MA distance branch, we outputs the MA distance branch only. In the post-processing, we utilized the thresholding to produce a binary gland region map, and gland markers. Then, the Watershed is used to produce the final segmentation. From the Table. 2, we noted that the designed multitask learning network outperform one task networks (Ours-INST, Ours-MA). Integrating both gland instance and MA distance map will produce a reliable performance in gland segmentation.

\begin{table}[h]
\small
\begin{center}
\begin{tabular}{|l|l|c|c|c|}
\hline
Datasets & Methods & F1(\%)$\uparrow$ & Obj-D(\%)$\uparrow$ & Obj-H$\downarrow$ \\
\hline\hline
GlaS & Ours-WoM & 90.2 & 89.8 & 54.4\\
                & TA-Net & \textbf{90.5} & \textbf{90.2} & \textbf{50.8} \\
\hline
CRAG & Ours-WoM & 83.7 & 89.1 & 108.6 \\
                & TA-Net & \textbf{84.2} & \textbf{89.3} & \textbf{105.2} \\
\hline
\end{tabular}
\end{center}
\caption{Ablation study on marker loss}
\vspace{-0.3em}
\end{table}

\begin{figure}[t]
  \begin{subfigure}[b]{0.24\linewidth}
    \includegraphics[width=\linewidth]{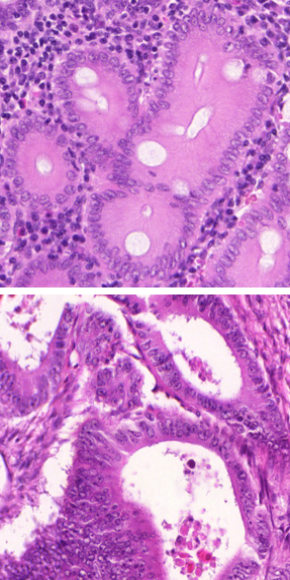}
     \caption{Images patch}
  \end{subfigure}
  \begin{subfigure}[b]{0.24\linewidth}
    \includegraphics[width=\linewidth]{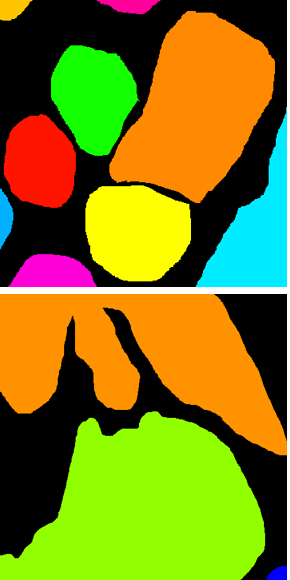}
     \caption{Ground truth}
  \end{subfigure}
  \begin{subfigure}[b]{0.24\linewidth}
    \includegraphics[width=\linewidth]{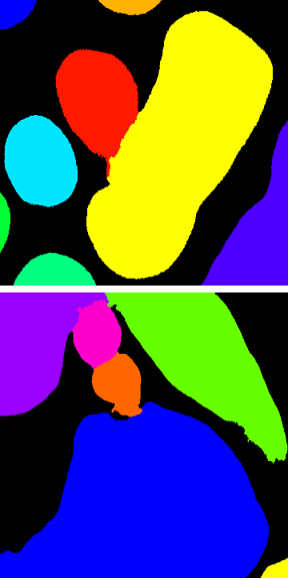}
     \caption{Ours-WoM}
  \end{subfigure}
  \begin{subfigure}[b]{0.24\linewidth}
    \includegraphics[width=\linewidth]{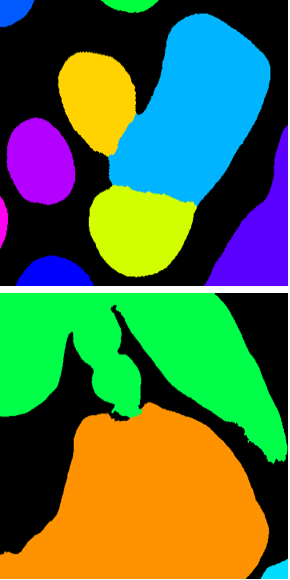}
     \caption{TA-Net}
  \end{subfigure}
   \caption{Examples of the effectiveness of the marker loss.}  
\label{fig:long}
\label{fig:onecol}
\vspace{-2em}
\end{figure}

\vspace*{3mm}
\textbf{W/ or w/o the marker loss.} The proposed TA-Net is compared with the same network without the marker loss (Ours-WoM). Table. 3. shows the results on two public datasets. From the quantitative results, we noted that the marker loss only improves the overall performance slightly on both two datasets. But we observed that the marker loss alleviates the over-segmentation and under-segmentation problems in clustered glands and deformed glands in many qualitative cases (Fig. 7). 

\begin{table}[t]
\small
\begin{center}
\begin{tabular}{|l|l|c|c|c|}
\hline
Datasets & Metrics & F1(\%)$\uparrow$ & Obj-D(\%)$\uparrow$ & Obj-H$\downarrow$ \\
\hline\hline
GlaS & Euclidean & 82.4 & 78.7 & 65.4 \\
                & Chessboard & 86.4 & 83.4 & 71.2 \\
                & MA & \textbf{90.2} & \textbf{89.8} & \textbf{54.4} \\
\hline
CRAG & Euclidean & 76.5 & 81.2 & 257.5 \\
                & Chessboard & 80.8 & 85.9 & 178.9 \\
                & MA & \textbf{83.7} & \textbf{89.1} & \textbf{108.6} \\
\hline
\end{tabular}
\end{center}
\caption{Ablation study on different distance metrics on CRAG dataset. }
\end{table}

\vspace*{3mm}
\textbf{Comparison on various distance map.} Deep Watershed-based regression algorithms provide the successful demonstration to separate the occluded objects and overlapped objects \cite{bai2017deep}. Most related method to ours is by Naylor \emph{et~al}. \cite{naylor2018segmentation}, which proposed a chessboard distance-based deep regression network for nuclei segmentation. First, comparing to their U-Net shape architecture, we employed the multitask learning-based Densely connected SegNet. We achieved promising performance in segmenting the gland foreground from the background. Second, the Medial Axis distance map preserve the topological property of the objects, which maintain the gland structure information during the training stage. Third, the marker loss will control the over-segmentation and under-segmentation issue for accurate marker detection. 

To demonstrate the effectiveness of MA distance transform with other distance-based metrics, we set up an experiment using our network test with different distance metrics, includes the Euclidean distance, Chessboard distance, Medial axis (MA) distance. To conduct a fair comparison, marker loss will not apply in this study, and we replace the MA distance metrics to other distance metrics. Similar to our post-processing approach, the predicted gland foreground segmentation and the predicted distance map are utilized to produce the final fine segmentation. From the results in Table. 4., we noted that our methods outperform the other two distance metrics. It gives the fact that medial axis distance metrics achieved the best performance comparing to use Euclidean and chessboard distance metrics.

\section{Conclusion }
In this paper, we propose a topology-aware network (TA-Net) to address the challenge of partitioning densely clustered glands in histopathology images. Firstly, the proposed multitask learning architecture integrates both instance segmentation and gland topology learning and learns their shared representation. Experimental results show that TA-Net outperforms the state-of-art multitask architectures, e.g., DCAN, and single-task architectures. Secondly, we propose a topology loss using Medial Axis distance map and gland makers. The loss penalizes the topology changes between the segmented glands and actual glands. The extensive experimental results on two public datasets demonstrate that the proposed TA-Net achieves state-of-the-art performance for densely clustered gland segmentation. In the future, we will extend the proposed approach to other challenging tasks, such as nuclei segmentation and semantic image segmentation.

\section*{Acknowledgement}
Research reported in this publication was supported by the National Institute of General Medical Sciences of the National Institutes of Health under Award Number P20GM104420. The content is solely the responsibility of the authors and does not necessarily represent the official views of the National Institutes of Health.

{\small
\bibliographystyle{ieee_fullname}
\bibliography{egbib}
}

\end{document}